\documentclass{article}
\usepackage{graphicx}
\usepackage{sectsty}
\usepackage{rotating}
\usepackage{subcaption}
\usepackage{amssymb}
\usepackage{ragged2e} 
\usepackage{authblk}
\usepackage[utf8]{inputenc}
\usepackage[margin=2cm]{geometry}
\usepackage{titlesec}
\usepackage{amsmath}
\usepackage{adjustbox}
\usepackage{tabularx}
\usepackage{multirow}
\usepackage{lipsum} 
\usepackage{url} 
\usepackage{geometry}
\usepackage{booktabs}
\usepackage{float}
\usepackage{amsmath}
\usepackage{algorithm}
\usepackage{algpseudocode}
\usepackage{svg}
\usepackage{comment}

\titleformat*{\section}{\large\bfseries}
\titleformat*{\subsection}{\normalsize\bfseries}
\titleformat*{\subsubsection}{\normalsize\bfseries}

\titleformat{\section}[hang]{\raggedright\bfseries}{\thesection}{1em}{}
\titleformat{\subsection}[hang]{\raggedright\bfseries}{\thesubsection}{1em}{}
\titleformat{\subsubsection}[hang]{\raggedright\bfseries}{\thesubsubsection}{1em}{}
\renewcommand\Affilfont{\small}

\title{\raggedright \justify \Large Pancreatic Tumor Segmentation as Anomaly Detection in CT Images Using Denoising Diffusion Models }

\author[a]{\raggedright Reza Babaei}
\author[a,*]{Samuel Cheng}
\author[b]{Theresa Thai}
\author[c]{Shangqing Zhao}

\affil[a]{\raggedright School of Electrical and Computer Engineering, University of Oklahoma, Norman, OK, United States  }
\affil[b]{\raggedright Department of Radiology, University of Oklahoma Health Sciences Center, Oklahoma City, OK , United Sates }
\affil[c]{\raggedright School of Computer Science, University of Oklahoma, Norman, OK, United States  }
\affil[*]{\raggedright Correspondence: Samuel Cheng , Email: samuel.cheng@ou.edu}

\date{}
\begin{document}
\maketitle

\renewcommand*{\Authsep}{, } 
\renewcommand*{\Authand}{, } 
\renewcommand*{\Authands}{, } 
\renewcommand*{\Affilfont}{\normalsize\itshape} 
\renewcommand*{\Authfont}{\normalsize}


\section*{Abstract}

Despite the advances in medicine, cancer has remained a formidable challenge. Particularly in the case of pancreatic tumors, characterized by their diversity and late diagnosis, early detection poses a significant challenge crucial for effective treatment. The intricate nature of abdominal anatomy further complicates identification, especially with small tumors, making it more challenging to discern diseased segments from healthy tissues. 

The advancement of deep learning techniques, particularly supervised algorithms, has significantly propelled pancreatic tumor detection in the medical field. However, supervised deep learning approaches necessitate extensive labeled medical images for training, yet acquiring such annotations is both limited and costly. Conversely, weakly supervised anomaly detection methods, requiring only image-level annotations, have garnered interest. Existing methodologies predominantly hinge on generative adversarial networks (GANs) or autoencoder models, which can pose complexity in training and, these models may face difficulties in accurately preserving fine image details.

This research presents a novel approach to pancreatic tumor detection, employing weak supervision anomaly detection through denoising diffusion algorithms. By incorporating a deterministic iterative process of adding and removing noise along with classifier guidance, the method enables seamless translation of images between diseased and healthy subjects, resulting in detailed anomaly maps without requiring complex training protocols and segmentation masks. This study explores denoising diffusion models as a recent advancement over traditional generative models like GANs, contributing to the field of pancreatic tumor detection. Recognizing the low survival rates of pancreatic cancer, this study emphasizes the need for continued research to leverage diffusion models' efficiency in medical segmentation tasks.

\section{Introduction}

The diversity of cancerous tumors and their late diagnosis, particularly in pancreatic cancer, make early detection difficult even with advanced technologies. The intricate abdominal area further complicates the identification of cancerous regions, especially in small pancreatic tumors. Medical imaging, particularly computed tomography (CT) scans, plays a vital role in detecting and segmenting pancreatic tumors. Despite the utility of medical images, distinguishing healthy and diseased regions within the pancreatic area based on image characteristics and intensities requires advanced expertise. Furthermore, manual tumor segmentation is labor-intensive and susceptible to errors. 

Recent advancements in deep learning models have shown reasonable performance in segmenting and classifying pancreatic tumors, involving stages such as image segmentation, feature extraction, and image classification \cite{iwasa2021automatic, lakkshmanan2022automated}. The representation of expanded pancreatic ducts in CT images has shown promise in facilitating early diagnosis, surgical preparation, and prognosis, as noted by Zou's study \cite{zou2023ctg}. Deep learning models like U-Net have exhibited promising outcomes in automatically segmenting pancreatic tumors from CT images \cite{lakkshmanan2022automated}. These supervised models hold promise for enhancing diagnostic accuracy by capturing enhancement patterns over time within the segmented area \cite{iwasa2021automatic}. 

The predominant focus of literature lies in supervised deep learning models, particularly those reliant on extensive datasets such as pixel-level segmentation masks. While datasets for pancreatic tumor segmentation have eased the validation of machine learning approaches in medical imaging \cite{xia2020synthesize}, the use of Generative Adversarial Networks (GANs) to produce synthetic medical images has also emerged to enhance limited datasets, thereby enhancing the generalization and efficacy of segmentation models \cite{anwar2018medical}, while facing challenges such as ensuring data quality and diversity, addressing domain adaptation issues, coping with limited training data, defining appropriate evaluation metrics, ensuring clinical relevance, and navigating ethical and legal considerations \cite{skandarani2023gans}. Ultimately, acquiring pixel-wise annotated ground truth in medical image analysis is notably challenging, often unavailable and limited, and prone to biases from human annotators\cite{suman2021quality}. 
Consequently, weakly supervised anomaly detection has garnered significant interest in research as a means to address these challenges. In contrast to fully supervised methods, image-level labels are only required for weakly supervised models to train the model.

 A significant innovation of this study lies in its pioneering use of denoising diffusion models, inspired by Wolleb's work \cite{wolleb2022diffusion}, to address the intricate task of segmenting pancreatic tumors in CT images. The application of denoising diffusion models has demonstrated their efficacy in enhancing CT images by reducing noise, thereby enhancing the quality of input data for segmentation algorithms and ultimately improving segmentation accuracy \cite{modaresi2023multi}. Utilizing a weakly supervised anomaly detection strategy, this approach depends solely on the image data and its associated image-level label, eliminating the segmentation map requirement during the model training process. During the training process, Denoising Diffusion Probabilistic Models (DDPM) and a binary classifier are trained using a dataset comprising samples from both healthy and diseased subjects. 

Through the encoding and denoising process, a deterministic sampling method outlined in Denoising Diffusion Implicit Models (DDIM), along with classifier guidance, effectively preserves tissues unaffected by disease while accurately representing diseased areas with realistic tissue depiction. 
By systematically simplifying the problem, this study reveals the latent capabilities of diffusion models and advocates for continued research efforts aimed at refining and optimizing these models for enhanced medical anomaly detection in pancreatic tumors.

\section{Related Works}

In the conventional paradigm of anomaly detection, autoencoders \cite{kingma2019introduction,zhou2017anomaly} has been specifically adapted for unsupervised anomaly detection within the domain of medical imaging \cite{marimont2021anomaly,zimmerer2018context}, typically trained on datasets comprising information from healthy subjects. Here, elevated anomaly scores flag any deviations from the established distribution, with the difference between the reconstructed healthy image and the anomalous input image serving as the basis for identifying anomalous pixels. However, a shift in focus has been observed in some methodologies towards GANs \cite{goodfellow2020generative} for tasks such as image-to-image translation \cite{siddiquee2019learning,wolleb2020descargan}. Nonetheless, the training of GANs poses considerable challenges, necessitating meticulous hyperparameter tuning and often requiring additional loss terms and architectural refinements to ensure consistent outcomes.

Recently, DDPMs have garnered attention for surpassing GANs in image synthesis \cite{dhariwal2021diffusion}. Expanding upon this success, DDPMs have been applied to various tasks including image reconstruction \cite{saharia2022palette}, registration \cite{kim2021diffusemorph}, segmentation \cite{baranchuk2021label}, and image-to-image translation \cite{choi2021ilvr,sasaki2021unit}. Particularly noteworthy are DDPMs discussed by Song \cite{song2020denoising} and Pinaya \cite{pinaya2022fast}, which have shown the capability to generate high-quality samples comparable to those produced by GANs, without the need for adversarial training. These models have achieved competitive log-likelihoods with transformers and maintained rapid inference times. 
Furthermore, the efficacy of diffusion probabilistic models has been explored in ``MedSegDiff," demonstrating their effectiveness in various vision tasks including image deblurring, super-resolution, and anomaly detection, thus highlighting their potential for medical image segmentation \cite{wu2022medsegdiff}. DDIMs \cite{song2020denoising}, closely related to score-based generative models \cite{song2020score}, have been employed for image interpolation and, more recently, for anomaly detection \cite{wolleb2022diffusion}. 

Recent advancements in denoising diffusion models hold significant potential to impact medical anomaly detection by enhancing input data quality, fortifying segmentation algorithm robustness, and ultimately improving the accuracy and efficiency of anomaly detection in medical images \cite{kazerouni2023diffusion}. Wolleb's research \cite{wolleb2022diffusion} presents an innovative semantic segmentation technique that harnesses diffusion models, improving lesion segmentation through adjustments in training and sampling strategies. Similarly, Behrend's work \cite{behrendt2024patched} proposes a patch-based estimation approach using diffusion models for brain anatomy reconstruction, achieving substantial relative improvement in tumor and multiple sclerosis lesion contexts.

These studies collectively advance the field of diffusion models in medical image analysis, showcasing their adaptability and effectiveness across various tasks such as segmentation, reconstruction, and anomaly detection, as highlighted by their diverse applications. However, the focus of these investigations has been on the relatively less complex task of brain tumor segmentation. It would be intriguing to explore how these techniques might be adapted to the more daunting challenge of pancreatic tumor segmentation, where the tumors are significantly less distinguishable from normal tissue.

\section{Method}

Considering the complex characteristics of pancreatic tumors, this study concentrated on refining the information prior to its processing by the diffusion model, aiming to improve its performance. Initial efforts centered on meticulously pre-processing the input CT images, leveraging the Medical Segmentation Decathlon (MSD) dataset and the data collected from the University of Oklahoma Health Sciences Center (OUHSC). The upcoming sections will provide details on each pre-processing step, provide a mathematical illustration of the diffusion model algorithm, and discuss the classifier guidance utilized in this investigation respectively.

\subsection{Pre-processing}

The initial pre-processing step involved centering and cropping the MSD pancreas dataset images to 128*128 dimensions, focusing on the pancreas heads and reducing model complexity, a key objective of this study. Regarding the existing labels, a decision was made to prioritize the pancreas head region over the entire pancreas. This decision was based on medical reasoning, aiming to optimize task performance given the rarity of tumors in the pancreas tail compared to those in the head. Consequently, the pancreas masks were used for positioning pancreas on the center of the image and cropping the CT images with desired dimensions to include only the head of the pancreas and, if present, any pancreatic tumors.

Following this, the subsequent step aimed to refine the cropped pancreas region by removing its surrounding tissues. This refinement process aimed to enhance the clarity of the pancreas's margins, emphasizing the task of distinguishing tumor from a healthy pancreas tissue rather than segmenting the pancreas itself. Figure 1(a) illustrates the isolated pancreas, after the surrounding tissues were removed, while Figure 1(b) displays its corresponding segmentation mask, which the segmentation mask was only used for resulting anomaly map evaluation stage. Figure 1(a) illustrates that the tumor boundary is not distinctly visible from the surrounding healthy pancreatic tissue, highlighting the challenges of pancreatic anomaly detection, as even radiologists face difficulties in accurately defining tumor margins, evidenced by the tumor shape presented in Figure 1(b).

Following the recommended dataset protocols, CT image normalization was executed, then ensuring all intensities were rescaled to a range between zero and one for optimal training efficacy. Additionally, to enhance model discernment between healthy and tumorous regions, slices exhibiting a mean intensity deviation of less than 0.1 between the healthy and non-healthy region were excluded from the training set. This meticulous processing underscores the precision and reliability of our training data, crucial for robust model performance in distinguishing between tumorous and healthy pancreatic tissue.

In contrast to the MSD dataset, the OUHSC test set does not provide ground truth for the pancreas segmentation mask. The data underwent similar pre-processing steps, but with a notable difference. Due to the absence of segmentation masks, the TotalSegmentator \cite{wasserthal2023totalsegmentator,isensee2021nnu} module was employed to segment the pancreas region in the images, not including its tumor mask. Subsequently, the slices were centered and cropped to match the input dimensions of the model, excluding surrounding tissues other than the pancreas, and finally intensity normalization was utilized to have values between zeros and ones. 

\begin{figure}[ht]
    \centering
    \begin{subfigure}[b]{0.3\textwidth} 
        \centering
        \includegraphics[width=\textwidth]{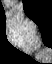}
        \caption{}
        \label{subfig:original_image}
    \end{subfigure}
    \hspace{0.15\textwidth} 
    \begin{subfigure}[b]{0.3\textwidth} 
        \centering
        \includegraphics[width=\textwidth]{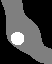}
        \caption{}
        \label{subfig:segmentation_mask}
    \end{subfigure}
    \caption{ (a) Subject's CT image and (b) its corresponding segmentation mask from the MSD dataset, after pre-processing steps}
    \label{fig1}
\end{figure}

\subsection{Anomaly Detection}
The key idea behind performing anomaly detection using a denoising diffusion model is to apply a trained diffusion model to transform an input image to resemble an image of a healthy subject. If the input image is already from a healthy subject, the output image will ideally show minimal changes. However, if the image is from an unhealthy subject, the changes will be concentrated near the anomaly, enabling effective anomaly detection.

Our approach primarily leverages a variant of the Denoising Diffusion Probabilistic Model (DDPM) \cite{ho2020denoising}, known as the Denoising Diffusion Implicit Model (DDIM) \cite{song2020denoising}, along with classifier guidance \cite{dhariwal2021diffusion}. In the following, we will more provide more details for each components. Since understanding DDPM is crucial for comprehending DDIM, we have included a brief introduction to DDPM for completeness.

\subsubsection{Denoising Diffusion Probabilistic Models}

A denoising diffusion model typically involves two phases: a forward diffusion process and a reverse denoising process. During the forward phase, noise is incrementally added to the input image $x$ through a series of steps, producing a sequence of progressively noisier images $\{\underbrace{x_0}_{x}, x_1, \ldots, x_N\}$, with $x_N$ approaching pure noise. This transformation is generally structured so that the noise added at each step is independent and Gaussian, simplifying the modeling process.

In the reverse denoising process, noise is gradually removed from the noisy image $x_N$ to recover the original image $x$. This reverse process is often accomplished using a deep learning model such as a U-Net \cite{ronneberger2015u}.

Mathematically, the forward process from $x_{n-1}$ to $x_n$ is characterized with conditional probability function
\begin{align}
q(x_n | x_{n-1}) = \mathcal{N}(x_n; \sqrt{1 - \beta_n} x_{n-1}, \beta_n I), 
\label{eqn:cond_prob}
\end{align}
where $\beta_n$ is a controlled parameter that determines the noise level at the $n$-the diffusion step. 
As we apply \eqref{eqn:cond_prob} recursively, we can readily obtain
\begin{align}
x_n = \sqrt{\bar{\alpha}_{n}} x_0 +  \sqrt{1 - \bar{\alpha}_{n}} \epsilon, \quad \text{with } \epsilon \sim \mathcal{N}(0,I). \quad 
\label{eqn:xn}
\end{align}
where  $\bar{\alpha}_n \triangleq \prod_{s=1}^{t} \alpha_s$. 
and 
$\alpha_n
\triangleq 1 - \beta_n$ Note that \eqref{eqn:xn} implies that instead of adding noise incrementally over $n$ steps to obtain the $n$-th noisy image $x_n$, we can attain $x_n$ by adding Gaussian noise only once. This simplification significantly eases the training process.

Let $\theta$ represent the model parameters involved in the reverse process. Our goal is to determine $\theta$ such that the model evidence is maximized, or equivalently, the negative log evidence $-\log p_\theta(x_0)$ is minimized. However, this objective is intractable. Consequently, we often aim to minimize the evidence lower bound (ELBO) as shown on the right-hand side of \eqref{eqn:elbo} instead:
\begin{align}
-\log(p_\theta(x_0)) \le E_{\sim q(x_1^N|x_0)}\left[\log \frac{q(x_1^N|x_0)}{p_\theta(x^N)}\right],
\label{eqn:elbo}
\end{align}
where 
$x_i^j$ is a shorthand notation for $\{x_i, x_{i+1}, \ldots, x_j\}$. Additionally, when $i=0$, we use $x_0^j=x^j$ to simplify the notation and reduce symbol clutter.

After some derivation, the ELBO on the right-hand side of \eqref{eqn:elbo} can be rewritten as follows:
\begin{align}
KL(q(x_N|x_0)||p(x_N))- E_{\sim q(x_1|x_0)}[\log{p_\theta(x_0|x_1)]}+\sum_{n=2}^N KL({q(x_{n-1}|x_{n},{ x_0})}||{p_\theta(x_{n-1}|x_n)}),
\label{eqn:elbo2}
\end{align}
where $KL(p(x) \|q(x))$ denotes the KL-divergence between two distributions $p(x)$.
 Note that the first term of \eqref{eqn:elbo2} does not depend on $\theta$ and hence can be safely discarded. 
 Let's examine the transition from $x_n$ to $x_{n-1}$ governed by $p_\theta(x_{n-1}|x_n)$ in the reverse process. This transition is influenced solely by the term $KL(q(x_{n-1}\|x_n, x_0) | p_\theta(x_{n-1}|x_n))$ in the third term. Using Bayes' rule, we can express $q(x_{n-1}|x_n,x_0)$ in a closed form as
\begin{align}
q(x_{n-1}|x_n,x_0)=\mathcal{N}(x_{n-1};\tilde{\mu}_n(x_n,x_0),\tilde{\beta}_n I)   
\end{align}
with $\tilde{\mu}_{n}({x}_{n},{x}_{0})\triangleq\frac{\sqrt{\bar\alpha_{n-1}}\beta_{n}}{1-\bar{\alpha}_{n}}{x}_{0}+\frac{\sqrt{\alpha_{n}}(1-\bar{\alpha}_{n-1})}{1-\bar{\alpha}_{n}}{x}_{n}
$ and $\tilde{\beta}_{n} \triangleq \,\frac{1-\bar{\alpha}_{n-1}}{1-\bar{\alpha}_{n}}\beta_{n}$. 

When the diffusion step is sufficiently small, the reverse process can be well-modeled as Gaussian also. So we can write $p_\theta(x_{n-1}|x_n)=\mathcal{N}(x_{n-1};\mu_\theta(x_n,n),\Sigma_\theta(x_n,n))$. Pick $\Sigma_\theta(x_n,n)$ as 
$\tilde \beta_n I$ such that  
$KL(q(x_{n-1}|x_{n},{ x_0})\| p_\theta(x_{n-1}|x_n))$ is minimize, we have
\begin{align}
    KL(q(x_{n-1}|x_{n},{ x_0}) \| p_\theta(x_{n-1}|x_n)) = \frac{1}{2 \tilde \beta_n^D} \left\|
\tilde{\mu}_n(x_n,x_0)-\mu_\theta(x_n,n)\right\|^2,
\end{align}
where $D$ is the dimension of $x$.
Apply \eqref{eqn:xn} to $\tilde{\mu}_{n}({x}_{n},{x}_{0})$, we have 
\begin{align}
    \tilde{\mu}_{n}({x}_{n},{x}_{0})=\frac{1}{\sqrt{\alpha_n}} \left(x_n - \frac{\beta_n}{\sqrt{1-\bar\alpha_n}}\epsilon \right)
    \label{eqn:mubar}
\end{align}
Reparametrize and write $\mu_\theta(x_n,n)$ with the same form as
\begin{align}
\mu_\theta(x_n,n)= \frac{1}{\sqrt{\alpha_n}} \left(x_n - \frac{\beta_n}{\sqrt{1-\bar\alpha_n}}\epsilon_\theta(x_n,n) \right),
\end{align}
we have 
\begin{align}
 KL(q(x_{n-1}|x_{n},{ x_0}) \| p_\theta(x_{n-1}|x_n)) &=\frac{1}{2 \tilde \beta_n^D} 
\left\|\frac{1}{\sqrt{\alpha_n}} \left(x_n - \frac{\beta_n}{\sqrt{1-\bar\alpha_n}}\epsilon \right)-\frac{1}{\sqrt{\alpha_n}} \left(x_n - \frac{\beta_n}{\sqrt{1-\bar\alpha_n}}\epsilon_\theta(x_n,n) \right)\right\|^2 \\
& =\frac{\beta_n^2}{2 \tilde \beta_n^D \alpha_n (1-\bar\alpha_n)} 
\left\|\epsilon -\epsilon_\theta(x_n,n) \right\|^2 
\end{align}





Consequently, the loss function of the reversed process (often modeled using a U-Net) can be set to


\begin{align}
L := \left\| \epsilon - \epsilon_{\theta}\left(x_n, n\right) \right\|^2,
\end{align}
where $x_n=\sqrt{\bar{\alpha}_n}x_0 + \sqrt{1 - \bar{\alpha}_n}\epsilon$ and $n$ are the input of the U-Net and $\epsilon_\theta(x_n,n)$ is the output trying to predict the Gaussian noise $\epsilon$.

Once the reverse model is trained, $x_{n-1}$ can be reconstructed as a sample of $\mathcal{N}(\tilde{\mu}_n(x_n,x_0),\tilde{\beta}_n I)$. Note that $\tilde{\mu}_n(x_n,x_0)$ as shown in \eqref{eqn:mubar} involved the unknown noise $\epsilon$, which can be predicted by the reverse model.  Thus, we finally have 
\begin{align}
x_{n-1} = \frac{1}{\sqrt{\alpha_n}} \left(x_n - \frac{\beta_n}{\sqrt{1-\bar\alpha_n}}\epsilon_\theta(x_n,n)  \right) +\sqrt {\tilde \beta_n }\epsilon 
\label{eqn:ddpm-reverse}
\end{align}
with $\epsilon \sim N(0,I)$.

\subsubsection{Denoising Defusion Implicit Model}

An inherent weakness of the DDPM model is that it requires many steps to produce a decent reconstruction, which can be very computationally expensive. Rather than assuming a Markovian structure, the Denoising Diffusion Implicit Model (DDIM) \cite{song2020denoising} allows for a non-Markovian structure. Consequently, significantly fewer iterations are needed to achieve performance comparable to that of DDPM.

Interestingly,  
it can be shown  \cite{song2020denoising} that  
the reverse step in DDPM 
as shown in \eqref{eqn:ddpm-reverse} 
can be rewritten as  
\begin{align}
x_{n-1} = \sqrt{\bar{\alpha}_{n-1}} \left( \frac{x_{n} - \sqrt{1 - \bar{\alpha}_n} \epsilon_{\theta}(x_{n}, n)}  {\sqrt{\bar{\alpha}_n} }\right) + \sqrt{\left(1 - \bar{\alpha}_{n-1}\right) - \sigma_n^2} \epsilon_{\theta}(x_{n}, n) + \sigma_n \epsilon
\quad 
\label{eqn:5}
\end{align}
with $\sigma_n = \sqrt{\tilde{\beta}_n}$.
In contrast, it turns out that \eqref{eqn:5} still applies to DDIM even when the Markovian assumption is dropped. However, we should set $\sigma_n$ to 0 in \eqref{eqn:5}, resulting in a deterministic sampling process. As elucidated in \cite{song2020denoising}, equation \eqref{eqn:5} can be likened to the Euler method for solving an ordinary differential equation (ODE). Consequently, the generation process can be reversed by employing the reversed ODE. Adequate discretization steps facilitate the encoding of $x_{n+1}$ given $x_n$ with
\begin{align}
x_{n+1} = x_n + \sqrt{\bar{\alpha}_{n+1}} \left( \sqrt{\frac{1}{\bar{\alpha}_{n}}} - \sqrt{\frac{1}{\bar{\alpha}_{n+1}}} \right) x_n + \left( \sqrt{\frac{1}{\bar{\alpha}_{n+1}} - 1} - \sqrt{\frac{1}{\bar{\alpha}_{n}} - 1} \right) \epsilon_{\theta}(x_n,n).
\label{eqn:6}
\end{align}

Utilizing equation \eqref{eqn:6} for $n \in \{0, \ldots, N-1\}$ facilitates the encoding of an image $x_0$ within a noisy image $x_N$. Subsequently, the original $x_0$ can be retrieved from $x_N$ using \eqref{eqn:5} with $\sigma_n = 0$ for $n \in \{N,...,1\}$.

\subsubsection{Classifier Guidance}

Ideally, it might be sufficient to train a diffusion model solely on healthy subjects, allowing it to convert any input into a representation of a healthy subject. However, in practice, the diffusion model would be more effective if it were also exposed to some unhealthy cases.

Inspired by GANs, we may train a classifier based on noisy images to guide the sampling process. Let $p_C(c|x,n)$ be the classifier output probability of class $c$ given input $x$, knowing $n$ forward steps have been applied. By leveraging the connection between diffusion models and score matching \cite{song2019generative}, we can incorporate the classifier's influence to revise the estimate of $\epsilon_\theta(x_n, n)$ to $\epsilon_\theta(x_n, n) - \sqrt{1 - \bar{\alpha}_n} \nabla_x \log p_C(h|x_n, n)$ \cite{dhariwal2021diffusion}. In practice, a gradient scale $S$ is introduced to increase the flexibility of the model (see Line 7 of Algorithm 1).

\subsubsection{Overall Algorithm}

We use images from both healthy and unhealthy subjects to train the DDIM model, where the training process is identical to that of the DDPM model. Additionally, we train a classifier that can determine whether an image is healthy or unhealthy, given the number of forward steps $n$.

Once the diffusion model and the classifier are trained, given an input image $x$ and a number of steps $N$, we apply \eqref{eqn:6} iteratively for $n \in \{0, \cdots, N-1\}$, resulting in a noise-like image $x_N$. Then, given the gradient scale $S$, we apply the reverse step as described in \eqref{eqn:5}, incorporating modifications on the noise estimate using the classifier bias adjustment. The anomaly is estimated by comparing the reconstructed image $x_0$ with the input image $x$. For simplicity, the net anomaly map is computed as the sum of the per-channel anomaly maps, as shown in Line 10 of Algorithm 1.

\begin{algorithm}
\caption{Detecting anomalies through noise encoding-decoding and classifier guidance}
\label{algorithm1}
\begin{algorithmic}[1]
\State \textbf{Input:} Original image $x$, label for the healthy class $h$, scale of the gradient $S$, noise level $N$
\State \textbf{Output:} Synthetic image $x_0$, Anomaly map $a$
\For{$t$ from $0$ to $N-1$}
    \State $x_{n+1} \gets x_n + \sqrt{\bar{\alpha}_{n+1}} \left[ \left(\sqrt{\frac{1}{\bar{\alpha}_n}} - \sqrt{\frac{1}{\bar{\alpha}_{n+1}}}\right) x_n + \left(\sqrt{\frac{1}{\bar{\alpha}_{n+1}} - 1} - \sqrt{\frac{1}{\bar{\alpha}_n} - 1}\right) \epsilon_{\theta}(x_n,n)\right]$
\EndFor
\For{$n$ from $N$ to $1$}
    \State $\epsilon' \gets \epsilon_{\theta}(x_n,n) - S\sqrt{1-\bar{\alpha}_n}\nabla_x\log p_C(h|x_n,n)$
    \State $x_{n-1} \gets \sqrt{\bar{\alpha}_{n-1}} \left(\frac{x_n - \sqrt{1-\bar{\alpha}_n}\epsilon'}{\sqrt{\bar{\alpha}_n}}\right) + \sqrt{1 - \bar{\alpha}_{n-1}} \epsilon'$
\EndFor
\State $a \gets \sum_{\text{channels}}  {|x - x_0|} $
\State \textbf{return} $x_0, a$
\end{algorithmic}
\end{algorithm}

\section{Experiments}

The resulting dataset, from the pre-processing steps, was partitioned into a 90-10\% split for training and testing respectively. This partitioning decision was driven by the limited data available and the requirement of the model to have sufficient training data samples while avoiding over-fitting. The classifier, the encoder block of the U-net model, underwent training for 10,000 iterations, achieving an overall accuracy of 79.08\% on the training set. The DDPM model was trained following the methodology outlined in \cite{nichol2021improved}, without incorporating data augmentation techniques. Hyperparameters for the DDPM model were selected according to the specifications provided in the appendix of \cite{dhariwal2021diffusion}, specifically configured to our case, for $T = 1000$ sampling steps.

\subsection{Dataset}

The MSD dataset encompasses various tasks related to medical segmentation, including cases focusing on small targets like pancreas tumors, underscoring the dataset's versatility and adaptability \cite{shen2021multi, jiang2022apaunet}. This dataset serves as a foundational benchmark for evaluating various deep learning models in pancreas segmentation, with researchers consistently assessing their performance against state-of-the-art methods \cite{oktay1804attention}, even benchmarking models like the U-Net on this specific task \cite{yang2022ax}. 

Through diverse studies, the MSD dataset has facilitated the evaluation of different segmentation models\cite{yan2021multi}, and several studies leveraging the MSD dataset have showcased the efficacy of deep learning methods for automated pancreas segmentation on a large scale \cite{lim2022automated}. Federated learning approaches have also found application with the MSD dataset for heterogeneous pancreas segmentation tasks, further illustrating its suitability for accommodating different segmentation methodologies \cite{liu2024pancreas}.

In this study, the training set images from the MSD dataset involved slices with tumor presence as non-healthy, while those displaying solely pancreatic tissue were classified as healthy. While acknowledging limitations in terms of dataset size and quality for pancreatic tumor segmentation \cite{suman2021quality}, the MSD dataset remains one of the most widely used resources in deep learning research. 

Besides the MSD dataset, we collected several clinical anonymous medical images from the OUHSC, solely for testing purposes. This test set medical images comprises CT images from various anonymous patients, which does not require human consent, and access to the data requires an official request. Unlike the MSD dataset, these images lack segmentation maps. However, they do contain tumor core locations cross marked in the relevant slices, which serve as the sole information available for the model evaluation. 

\section{Results}

The utilized model has two primary hyperparameters: the classifier guidance scale (S) and the noise level added to the inputs (N). These parameters play a crucial role in denoising images from unhealthy to healthy subjects, while incorporating feedback from the classifier to adjust gradients.

The guidance scale required for matching textures varied depending on the discernibility of healthy and non-healthy regions. For instance, in cases such as pancreatic cancer where textures appear similar at first glance, lower guidance is recommended to avoid over emphasising gradient paradigms. Conversely, when significant differences exist between healthy and non-healthy regions, higher guidance is necessary to facilitate accurate conversion between classes. Through experimentation, optimal guidance scales were found to range between 5-10, with a scale of 10 being slightly more aggressive. Similarly, if the noise level N is selected too large, it leads to image distortion. Conversely, if N is chosen too small, the model lacks the necessary flexibility to accurately remove the tumor from the image. In this study, optimal N was determined to be most effective within the range of 200-300 for optimal denoising performance. 

To facilitate quantitative comparison, anomaly maps were normalized, and a threshold was applied to generate binary maps for Dice score calculation. Lower thresholds resulted in larger tumorous regions, compromising visual clarity, while higher thresholds led to the loss of valuable information and potential tumor locations. Varying thresholds were tested, with Figure 3 demonstrating the effect of applying a 35\% threshold to anomaly heat maps showcased in Figure 2. Given the classifier's imperfect ability to distinguish healthy and non-healthy regions, misclassified cases were excluded from the Dice score calculation. Results of the Dice score calculations with different hyperparameter configurations are presented in Tables 1 and 2.

\begin{figure}[htbp]
    \centering
    \begin{subfigure}[b]{0.3\textwidth}
        \centering
        \includegraphics[width=\textwidth]{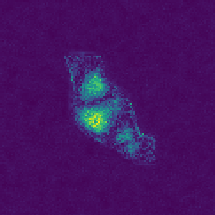} 
        \caption{S=5, N=200}
    \end{subfigure}
    \hfill
    \begin{subfigure}[b]{0.3\textwidth}
        \centering
        \includegraphics[width=\textwidth]{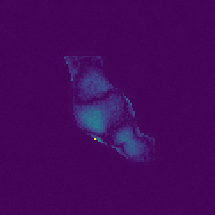} 
        \caption{S=5, N=300}
    \end{subfigure}
    \hfill
    \begin{subfigure}[b]{0.3\textwidth}
        \centering
        \includegraphics[width=\textwidth]{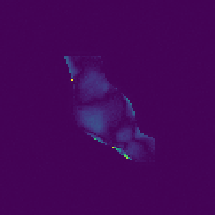} 
        \caption{S=5, N=400}
    \end{subfigure}
    \hfill
    \begin{subfigure}[b]{0.3\textwidth}
        \centering
        \includegraphics[width=\textwidth]{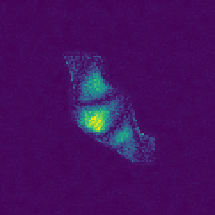} 
        \caption{S=8, N=200}
    \end{subfigure}
    \hfill
    \begin{subfigure}[b]{0.3\textwidth}
        \centering
        \includegraphics[width=\textwidth]{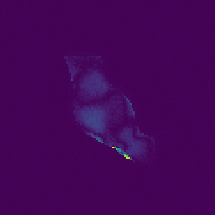} 
        \caption{S=8, N=300}
    \end{subfigure}
    \hfill
    \begin{subfigure}[b]{0.3\textwidth}
        \centering
        \includegraphics[width=\textwidth]{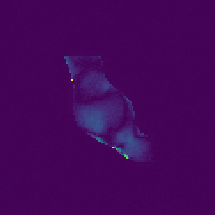} 
        \caption{S=8, N=400}
    \end{subfigure}
    \hfill
    \begin{subfigure}[b]{0.3\textwidth}
        \centering
        \includegraphics[width=\textwidth]{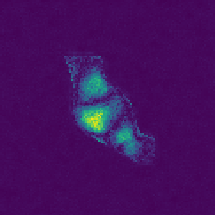} 
        \caption{S=10, N=200}
    \end{subfigure}
    \hfill
    \begin{subfigure}[b]{0.3\textwidth}
        \centering
        \includegraphics[width=\textwidth]{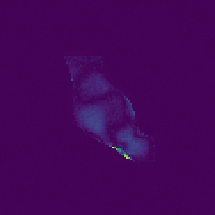} 
        \caption{S=10, N=300}
    \end{subfigure}
    \hfill
    \begin{subfigure}[b]{0.3\textwidth}
        \centering
        \includegraphics[width=\textwidth]{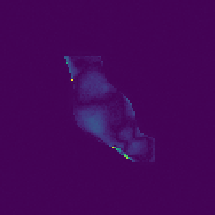} 
        \caption{S=10, N=400}
    \end{subfigure}
    \caption{Anomaly map of the subject, with different noise levels (N) and classifier guidance scales (S) }
    \label{fig2}
\end{figure}

\begin{figure}[htbp]
    \centering
    \begin{subfigure}[b]{0.3\textwidth}
        \centering
        \includegraphics[width=\textwidth]{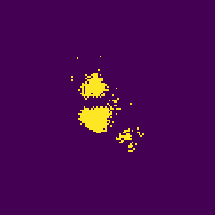} 
        \caption{S=5, N=200}
    \end{subfigure}
    \hfill
    \begin{subfigure}[b]{0.3\textwidth}
        \centering
        \includegraphics[width=\textwidth]{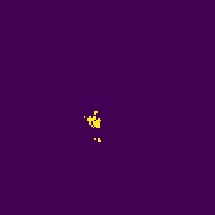} 
        \caption{S=5, N=300}
    \end{subfigure}
    \hfill
    \begin{subfigure}[b]{0.3\textwidth}
        \centering
        \includegraphics[width=\textwidth]{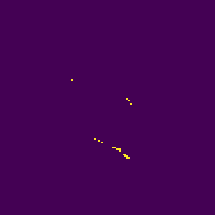} 
        \caption{S=5, N=400}
    \end{subfigure}
    \hfill
    \begin{subfigure}[b]{0.3\textwidth}
        \centering
        \includegraphics[width=\textwidth]{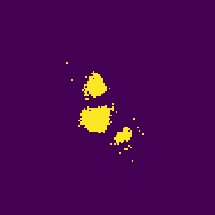} 
        \caption{S=8, N=200}
    \end{subfigure}
    \hfill
    \begin{subfigure}[b]{0.3\textwidth}
        \centering
        \includegraphics[width=\textwidth]{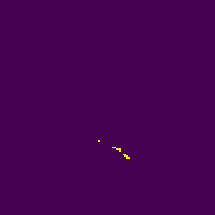} 
        \caption{S=8, N=300}
    \end{subfigure}
    \hfill
    \begin{subfigure}[b]{0.3\textwidth}
        \centering
        \includegraphics[width=\textwidth]{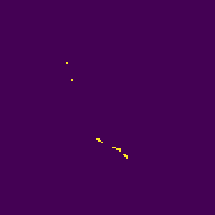} 
        \caption{S=8, N=400}
    \end{subfigure}
    \hfill
    \begin{subfigure}[b]{0.3\textwidth}
        \centering
        \includegraphics[width=\textwidth]{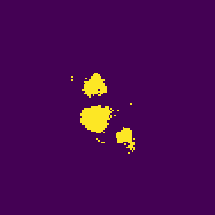} 
        \caption{S=10, N=200}
    \end{subfigure}
    \hfill
    \begin{subfigure}[b]{0.3\textwidth}
        \centering
        \includegraphics[width=\textwidth]{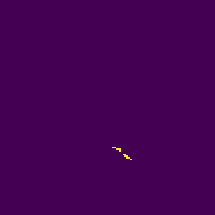} 
        \caption{S=10, N=300}
    \end{subfigure}
    \hfill
    \begin{subfigure}[b]{0.3\textwidth}
        \centering
        \includegraphics[width=\textwidth]{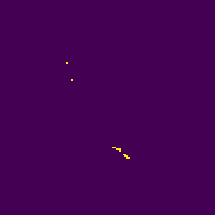} 
        \caption{S=10, N=400}
    \end{subfigure}
    \caption{Segmentation map of the subject, with different noise levels (N) and classifier guidance scales (S), and a fixed threshold of 35\% }
    \label{fig3}
\end{figure}

\begin{table}[h]
\centering
\begin{tabular}{cccc}
\toprule
\multirow{2}{*}{\textbf{Classifier Guidance Scale}} & \multirow{2}{*}{\textbf{Threshold}} & \multicolumn{2}{c}{\textbf{Performance Metrics}} \\ \cmidrule(l){3-4} 
 &  & \textbf{Dice Score} & \textbf{Classifier Accuracy (\%)} \\ \midrule
\multirow{5}{*}{5} & 25 & 0.320 & 82.75 \\
 & 30 & 0.267 & 81.60 \\
 & 35 & 0.214 & 79.31 \\
 & 40 & 0.188 & 81.60 \\
 & 45 & 0.162 & 82.75 \\ \midrule
\multirow{4}{*}{8} & 35 & 0.197 & 91.02 \\
 & 40 & 0.170 & 89.74 \\
 & 45 & 0.152 & 91.02 \\
 & 50 & 0.117 & 89.74 \\ \midrule
\multirow{4}{*}{10} & 35 & 0.218 & 91.02 \\
 & 40 & 0.157 & 89.74 \\
 & 45 & 0.141 & 91.02 \\
 & 50 & 0.123 & 92.30 \\ \bottomrule
\end{tabular}
\caption{Impact of different guidance scales and thresholds on performance metrics, with fixed noise level (N) of 300}
\label{table1}
\end{table}

\begin{table}[h]
\centering
\begin{tabular}{lccc}
\toprule
\textbf{Metric} & \textbf{Noise Level (N)} & \textbf{Dice Score} & \textbf{Classifier Accuracy (\%)} \\ \midrule
\multirow{3}{*}{\textbf{S 5, Th 35}} & 200 & 0.243 & 81.60 \\
 & 300 & 0.214 & 79.31 \\
 & 400 & 0.159 & 79.31 \\ \bottomrule
\end{tabular}
\caption{Impact of noise level (N) with fixed guidance scale (S) and threshold (Th) on performance metrics}
\label{table2}
\end{table}

In addition to evaluating the proposed diffusion model using the publicly available MSD dataset, we assessed the proposed method performance and generalization ability when applied to images not captured under the same environment with some anonymous clinical images collected from the OUHSC. This setup provided an opportunity to assess the diffusion model's performance with unlabeled data from real clinical scenarios. The same trained diffusion model weights used on the MSD dataset evaluation set, were applied to test the model's performance. A classifier guidance scale of 7 and noise level of 300 were found to be qualitatively optimal. For qualitative examination, the actual tumor locations from the OUHSC data were marked with crosses in the figures. Figure 4 displays anomaly maps and the associated classifier confidence levels for the example cases. Cases (a) and (b) were accurately classified by the model with high confidence, and the localization peaks align closely with the indicated cross marks. Conversely, case (c) demonstrates lower confidence in its classification, resulting in a localization peak that is distanced from the marked location. Since segmentation masks were unavailable for this test set images, only qualitative measures could be conducted. 

Distinguishing between healthy and non-healthy regions within the OUHSC test set may not be straightforward. Furthermore, the performance of the classifier lacks a high degree of accuracy. However, there is a tendency for the anomaly maps to be more accurately localized when the confidence level of the classifier is higher, which may provide useful indications of the trustworthiness of the candidate tumor location. 
The confidence levels of the test subjects in Figure 4 are included in the caption. 

Recognizing the valuable fact that diffusion models do not rely on segmentation masks for training, this study aims to initiate an ongoing exploration into utilizing denoising diffusion models for pancreatic tumor detection as anomaly detection, acknowledging that the findings may not match the state-of-the-art supervised algorithms documented in the literature.

\begin{figure}[htbp]
    \begin{subfigure}[b]{0.3\textwidth}
        \centering
        \includegraphics[width=\textwidth]{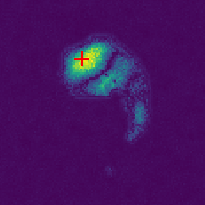} 
        \caption{ }
    \end{subfigure}
    \hfill
    \begin{subfigure}[b]{0.3\textwidth}
        \centering
        \includegraphics[width=\textwidth]{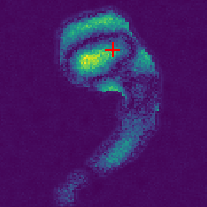} 
        \caption{ }
    \end{subfigure}
    \hfill
    \begin{subfigure}[b]{0.3\textwidth}
        \centering
        \includegraphics[width=\textwidth]{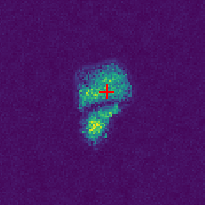} 
        \caption{}
    \end{subfigure}
    \caption{
    The anomaly heat map results of the OUHSC test set acquired from the diffusion model accompanied by the cross marked approximate tumor locations, and the corresponding classifier confidence levels of 0.9971, 0.9960, and 0.4645 respectively. }
    \label{fig4}
\end{figure}

\section{Discussion}

This study undertook an investigation into the diffusion models customized for medical image anomaly detection, with a specific focus on identifying pancreatic cancer. 
The preliminary work of this study shows that utilizing anomaly detection through denoising diffusion models, provides promising future directions to lower the dependence on challenging segmentation masks and use the potentials of the diffusion models. It is crucial to note that the method employed in this study is not reliant on pixel-level segmentation masks, thereby mitigating potential human biases during manual image segmentation, despite not achieving state-of-the-art results in terms of the highest dice scores. Potentially, numerous medical images without segmentation masks from different resources would come along in this effort, which have been not used in supervised training before, because of lacking segmentation masks. Given the more challenging acquisition of anomaly maps with the OUHSC test set, our findings suggest that diffusion models could represent a significant advancement in pancreatic tumor detection.

A pivotal aspect of the investigation revolved around the meticulous tuning of hyperparameters governing both the classifier guidance scale and the noise level incorporated into the model. This optimization process was crucial in striking a delicate balance, ensuring that the model achieved effective denoising of images while simultaneously preserving the image details. Results highlighted the intricate interplay between these parameters, emphasizing the need for careful calibration to achieve optimal performance. Our findings revealed that the performance of the classifier remained at a moderate level of accuracy. This observation underscores the inherent challenges associated with accurately distinguishing between healthy and non-healthy regions within complex medical imaging datasets, particularly in the context of pancreatic pathology. Looking ahead, future research directions may involve the exploration of alternative model architectures and the integration of multi-modal imaging data to further enhance anomaly detection and characterization. Furthermore, the subjective process of threshold selection for Dice score calculation introduces a level of uncertainty into the findings, underscoring the necessity for standardized methodologies in future research efforts.

\section{Conclusion}

Pancreatic cancer remains a formidable challenge despite advancements in medical imaging and the utilization of deep learning techniques. Although supervised learning has significantly progressed medical segmentation, the acquisition of extensive pixel-level labels remains costly. However, the emergence of diffusion models, exhibiting superior capabilities over GANs in image synthesis, presents an opportunity for enhancing medical segmentation through anomaly detection. While diffusion models have been applied mostly on brain tumor detection, this study is the first study that focuses on implementing medical anomaly detection specifically for pancreatic tumor detection through the development of solely denoising diffusion models, and further developing the potentials with modifications to the algorithm or the dataset, will be shifted to future works. 

\section*{Author contributions}

RB preprocessed the dataset, implemented algorithms, and drafted the manuscript. TT gathered test data from the OUHSC clinic. SC, TT and SZ provided guidance, reviewed and edited the manuscript, and contributed to result discussions. All authors contributed to the article and approved the final version for submission.

\section*{Informed Consent}

The medical images utilized in this study were either sourced from public databases, such as the MSD dataset, or anonymized, as was the case with the OUHSC test data, to safeguard privacy. Therefore, ethical review and individual consent were deemed unnecessary for this research.

\section*{Declaration of generative AI and AI-assisted technologies in the writing process}

During the preparation of this work the authors used ChatGPT in order to check for the content fluency and possible grammatical errors only. After using this tool/service, the authors reviewed and edited the content as needed and take full responsibility for the content of the publication.

\section*{Data Availability}

The MSD training dataset used in this study is available to public. The OUHSC test set availability is subjected to formal request. 

\section*{Conflicts of Interest}

The authors declare that the research was conducted in the absence of any commercial or financial relationships that could be construed as a potential conflict of interest.

\section*{Funding Statement}

This work is partially supported by the Vice President for Research and Partnerships of the University of Oklahoma, the Data Institute for Societal Challenges, and the Stephenson Cancer Center through DISC/SCC Seed Grant Award.

\bibliographystyle{unsrt}
\bibliography{bibliography}

\end{document}